\documentclass[sigconf, nonacm]{acmart}

\usepackage{nicefrac}
\usepackage{microtype}
\usepackage{subcaption}
\usepackage{multirow}
\usepackage{enumitem}
\usepackage{tikz}
\usetikzlibrary{shapes.geometric,arrows.meta,positioning,fit,backgrounds,calc,decorations.pathreplacing}
\usepackage{pgfplots}
\pgfplotsset{compat=1.18}

\definecolor{accentblue}{RGB}{30,90,160}
\definecolor{mutedgray}{RGB}{105,105,105}
\definecolor{profilecol}{RGB}{255,220,180}
\definecolor{activitycol}{RGB}{180,220,255}

\begin{document}

\title{Quantizing Intent: Cross-Domain Semantic IDs from Organic Activity for Industrial Ranking}

\author{Julie Choi}
\authornote{Corresponding author.}
\email{julchoi@linkedin.com}
\affiliation{%
  \institution{LinkedIn}
  \country{United States}
}

\author{Haoran Ye}
\authornotemark[1]
\email{haye@linkedin.com}
\affiliation{%
  \institution{LinkedIn}
  \country{United States}
}

\author{Zhiwei Ding}
\authornotemark[1]
\email{zding@linkedin.com}
\affiliation{%
  \institution{LinkedIn}
  \country{United States}
}

\author{Bo Long}
\email{BLong@linkedin.com}
\affiliation{%
  \institution{LinkedIn}
  \country{United States}
}

\author{Benjamin Zelditch}
\email{bzelditch@linkedin.com}
\affiliation{%
  \institution{LinkedIn}
  \country{United States}
}

\author{Arpita Vats}
\email{avats@linkedin.com}
\affiliation{%
  \institution{LinkedIn}
  \country{United States}
}

\begin{abstract}
Ads click-through rate (CTR) prediction is constrained by sparse user supervision: most users engage with ads infrequently while generating dense behavioral evidence in organic surfaces such as feed. Transferring these cross-domain signals into ads ranking is difficult due to domain mismatch, serving cost, and production complexity.

We introduce \textbf{cross-domain user Semantic IDs (SIDs)} derived from organic feed activity and establish that \textbf{behavioral activity richness governs cross-domain transfer quality}: SIDs from user profile text yield $+0.036\%$ AUC; SIDs from a LLaMA-3.1-based user embedding model (contrastively fine-tuned on cross-domain activity data, with user profile as the input prompt) yield $+0.107\%$; and SIDs from direct feed activity behavioral embeddings yield $+0.213\%$. This monotonic progression reveals that the amount of behavioral activity encoded in the source representation is the key determinant of downstream ads ranking quality.

We propose \textbf{RQ-FSQ (Residual + Finite Scalar Quantization)}, a method that pairs per-dimension scalar quantization with residual VAE quantization to discretize pre-trained embeddings while matching dense-embedding AUC. RQ-FSQ matches or slightly exceeds the AUC of the dense source embedding on two heterogeneous sources --- Feed Activity ($+0.351\%$ at $\sim 30\times$ smaller storage) and Activity-Tuned LLaMA ($+0.265\%$ at $\sim 280\times$ smaller storage) --- establishing it as a general-purpose discretizer for pre-trained embeddings. We further introduce the \textbf{Hierarchical Discrete Embedding (HDE) Module}, which encodes any $K$-level SID via prefix n-gram sparse embedding tables trained end-to-end under the CTR objective.

On a large-scale industrial production ads system, cold-start segment analysis shows gains reaching $+1.522\%$ for users with near-zero ad interaction history, directly validating the cross-domain behavioral transfer mechanism.
\end{abstract}

\ccsdesc[500]{Information systems~Recommender systems}
\ccsdesc[500]{Information systems~Online advertising}
\ccsdesc[300]{Computing methodologies~Neural networks}
\ccsdesc[300]{Computing methodologies~Learning latent representations}

\keywords{Semantic IDs, Cross-Domain Recommendation, Ads CTR Prediction, Residual Quantization, Finite Scalar Quantization, Generative Recommendation}

\maketitle

\section{Introduction}

Ads recommendation systems operate under extreme interaction sparsity. Compared with organic surfaces, where users generate frequent and diverse behavioral feedback, ad clicks are relatively rare and unevenly distributed across the user population. This creates a persistent challenge: for the large fraction of users with sparse ad interaction history, the model has little behavioral signal from which to infer intent. The problem is most acute in the \emph{cold-start} regime --- new or infrequent ad engagers --- where the model must estimate user preferences from profile attributes alone.

A natural source of supplemental evidence is organic feed activity. Users interact with feed content orders of magnitude more often than with ads, and those interactions encode evolving topical interests, content-type affinities, and engagement intent. However, transferring feed signals into an ads CTR model is difficult for three reasons: \textbf{(1)}~domain mismatch between feed and ads feature spaces; \textbf{(2)}~the high dimensionality and serving cost of dense behavioral embeddings; and \textbf{(3)}~integration complexity within large-scale production training pipelines that cannot accommodate bespoke preprocessing per feature source.

Semantic IDs (SIDs)~\citep{tiger2023,semidranking2023} have recently emerged as an effective way to discretize dense embeddings into compact token sequences via residual quantization (RQ-KMeans or RQ-VAE~\citep{rqvae2022}), enabling standard embedding-table lookup and end-to-end fine-tuning under the downstream task objective. Prior SID work focuses almost exclusively on \emph{single-domain} use: the embeddings being quantized originate in the same domain as the recommendation target~\citep{tiger2023,semidranking2023,idgenrec2024,lcrec2024,meta2025sid}. In this paper we study a different setting --- \emph{cross-domain viewer SIDs} --- where user representations derived from organic feed behavior are quantized and used as input features for ads CTR prediction. To our knowledge, this is the first empirical study of cross-domain viewer SIDs for industrial ads CTR modeling.

We integrate cross-domain viewer SIDs into a production decoder-only Transformer ranking model for ads CTR. The key design challenge is the embedding module: how to map a $K$-level discrete token sequence into a dense representation that is trainable end-to-end. We introduce the \textbf{Hierarchical Discrete Embedding (HDE) Module}, which encodes each SID via prefix n-gram sparse embedding tables --- prefix unigram, bigram, and trigram keys are each hashed to indices in dedicated trainable tables with bounded memory.

\paragraph{Contributions.}
\begin{itemize}[leftmargin=*]
  \item We propose \textbf{RQ-FSQ (Residual + Finite Scalar Quantization)}, a quantization method for pre-trained embeddings that pairs per-dimension scalar quantization (FSQ) with residual VAE quantization (RQ-VAE), preserving global geometry (RQ-VAE) and per-dimension fine structure (FSQ). RQ-FSQ matches or slightly exceeds dense-embedding AUC on two heterogeneous sources --- Feed Activity ($+0.351\%$ at $\sim 30\times$ smaller storage) and Activity-Tuned LLaMA ($+0.265\%$ at $\sim 280\times$ smaller storage) --- establishing it as a general-purpose discretizer for pre-trained embeddings.
  \item We introduce \textbf{cross-domain viewer SIDs} and establish the \textbf{behavioral activity richness principle}: the quality of cross-domain SID transfer scales monotonically with the amount of behavioral activity encoded in the source embedding --- profile text ($+0.036\%$), activity-trained embeddings ($+0.107\%$), and direct behavioral aggregations ($+0.213\%$), with cold-start segment gains reaching $+1.522\%$. This provides a general design principle for selecting SID sources.
  \item We introduce \textbf{Multi-Source SID}: a structured 9-code representation where $c_1$--$c_3$ come from Activity-Tuned LLaMA (backbone; highest user coverage), $c_4$--$c_6$ from Profile Qwen (missing codes inferred from Activity-Tuned LLaMA embeddings during SID training), and $c_7$--$c_9$ from Feed Activity. Multi-Source SID achieves $+0.036\%$ AUC over na\"{i}vely summing three independent single-source SIDs, with the HDE Module encoding each source's prefix n-gram tables independently and summing at the input layer --- no architectural changes required.
\end{itemize}

\section{Related Work}

\paragraph{Sequential recommendation and ads CTR.}
Transformer-based~\citep{vaswani2017attention} sequential architectures are now the dominant backbone for recommendation and ads CTR. SASRec~\citep{sasrec2018} applies left-to-right self-attention to user interaction sequences; BERT4Rec~\citep{bert4rec2019} extends this with bidirectional masked prediction. For ads CTR specifically, BST~\citep{bst2019} introduces behavior-sequence Transformers, while DIN~\citep{din2018} and DIEN~\citep{dien2019} model adaptive interest activation and evolution. DCN-V2~\citep{dcnv2_2021} combines explicit feature crossing with deep networks for large-scale ranking. The production baseline used in our experiments is a decoder-only Transformer for ads CTR with context-conditioned attention, timestamp-based RoPE, session masking, and FlashAttention~\citep{cadet2026,flashattn2022}. This paper treats that model as a fixed backbone and focuses entirely on improving viewer-side input representations.

\paragraph{Semantic IDs for recommendation.}
TIGER~\citep{tiger2023} is the seminal work on SIDs for recommendation: it quantizes item content embeddings via RQ-VAE~\citep{rqvae2022} into hierarchical codeword tuples and trains a seq2seq Transformer to autoregressively decode the SID of the next item. \citet{semidranking2023} extend SIDs to the YouTube ranking stage using SentencePiece tokenization over codebook sequences, demonstrating improved generalization on tail and new items in production. IDGenRec~\citep{idgenrec2024} aligns LLMs with recommendation via jointly learned textual item IDs; LC-Rec~\citep{lcrec2024} integrates collaborative filtering signals via vector quantization. LETTER~\citep{letter2024} proposes learnable tokenization that adapts codebook assignment jointly with the downstream recommendation objective. \citet{meta2025sid} deploy semantic ID prefix n-grams for item IDs in production ads ranking at Meta. \citet{side2025} introduce SIDE, which converts VQ codewords to collision-free discrete IDs via positional base-$C$ encoding, enabling parameter-efficient embedding table lookups for long sequences.

A common thread in all of these works is the \emph{single-domain} assumption: SIDs are derived from the same interaction space as the recommendation target. We break this assumption by constructing \emph{user} SIDs from cross-domain organic activity and using them for ads CTR prediction.

\paragraph{Cross-domain recommendation.}
Cross-domain recommendation has been studied through embedding-and-mapping (EMCDR~\citep{emcdr2017}), shared hidden layer activations (CoNet~\citep{conet2018}), and through discrete SID tokens for item vocabulary alignment across domains (GenCDR~\citep{gencdr2026}). These methods typically require overlapping item or user sets between domains, or explicit domain-adaptation objectives. Our approach uses discrete SID tokens as a \emph{transfer interface} with no explicit alignment loss: the CTR objective drives adaptation of the cross-domain embedding tables to the target task.

\paragraph{Discrete representation learning.}
VQ-VAE~\citep{vqvae2017} introduces vector quantization with straight-through gradients for discrete latent spaces. RQ-VAE~\citep{rqvae2022} extends this with residual quantization, enabling hierarchical multi-level codebooks where successive residuals capture progressively finer structure --- the foundation of SID hierarchies. We use RQ-KMeans (deterministic, no variational objective) for production stability. Our HDE Module (Section~\ref{sec:hde}) shares the goal of sparse trainable tables for discrete IDs with~\citet{side2025}, but adopts prefix n-gram \emph{hash-based} lookups with bounded memory $H^{\max}$, in contrast to SIDE's collision-free positional base-$C$ encoding which scales as $O(C^K)$. This trade-off --- controlled collision for bounded memory --- is required for industrial-scale deployment where unbounded tables are infeasible. We further extend the design to a Multi-Source SID setting where prefix tables are applied per-source independently (Section~\ref{sec:multisource-design}).

\section{Production Ads Ranking Model}
\label{sec:baseline}

The baseline system is a decoder-only Transformer~\citep{cadet2026} trained for industrial ads CTR prediction. Given a user's interaction sequence $\mathbf{x} = (e_1, \ldots, e_L)$ over a trailing window, the model predicts click probability autoregressively. Each event $e_t$ carries embeddings for ad creative ID, campaign ID, format and charge type, OS, objective type, and other contextual features. Context-conditioned attention with timestamp RoPE handles positional information; session masking aligns the attention pattern at training with the causal structure at serving time. Training uses DDP/FSDP2~\citep{fsdp2023} on H200 clusters. This paper extends the baseline with cross-domain viewer SIDs while leaving the Transformer backbone, attention stack, and prediction head \textbf{unchanged}: all improvements come from richer user input representations.

\section{Method}

\subsection{SID Construction: RQ-KMeans and RQ-FSQ}
\label{sec:sid-gen}

For a dense user embedding $\mathbf{v} \in \mathbb{R}^d$, residual quantization produces a $K$-tuple of discrete tokens:
\begin{equation}
  \mathrm{SID}(\mathbf{v}) = (c_1, c_2, \ldots, c_K), \quad c_k \in \{1, \ldots, C\},
  \label{eq:sid}
\end{equation}
where $C$ is the codebook size and $K$ is the number of quantization levels. We use two quantization methods, chosen based on whether the SID is trained \emph{from scratch} or must \emph{align with a pre-trained embedding}:

\paragraph{RQ-KMeans.} Operates deterministically over $K$ residual stages: (1) assign $\mathbf{v}$ to its nearest centroid in a $C$-entry codebook, recording $c_1$; (2) compute residual $\mathbf{r}_1 = \mathbf{v} - \boldsymbol{\mu}_{c_1}$; (3) repeat on successive residuals to obtain $c_2, \ldots, c_K$. Because the uniform-variance assumption of $k$-means fits well when quantizing embeddings from scratch, RQ-KMeans achieves excellent codebook utilization and produces reproducible offline assignments.

\paragraph{RQ-FSQ (Residual + Finite Scalar Quantization).}\label{fsqrq} A quantization method for pre-trained embeddings that preserves information at two complementary scales: global geometry (via RQ-VAE) and per-dimension fine structure (via FSQ). FSQ~\citep{mentzer2023fsq} independently quantizes each dimension of $\mathbf{v}$ to a finite integer alphabet $\mathcal{L} = \{-L, \ldots, L\}$:
\begin{equation}
  \hat{\mathbf{C}}_{\mathrm{FSQ}} = \mathrm{FSQ}(\mathbf{v}) = \mathrm{round}\!\left(\tanh(\mathbf{v}) \cdot L\right).
  \label{eq:fsq}
\end{equation}
RQ-VAE~\citep{rqvae2022} quantizes successive residuals across $K$ stages, yielding $\hat{\mathbf{C}}_{\mathrm{RQ}} = (c_1, \ldots, c_K)$ where $c_i$ is the nearest codebook entry at stage $i$. The two streams are fused additively in the downstream model:
\begin{equation}
\mathbf{e}^{\text{RQ-FSQ}} = \mathbf{e}^{\hat{\mathbf{C}}_{\mathrm{RQ}}} + f(\hat{\mathbf{C}}_{\mathrm{FSQ}}),
\label{eq:rqfsq-fuse}
\end{equation}
where $\mathbf{e}^{\hat{\mathbf{C}}_{\mathrm{RQ}}}$ is the downstream embedding for the RQ codes and $f$ is a linear projection to the model dimension. We apply RQ-FSQ to the Feed Activity source (Section~\ref{sec:fsq}).

\paragraph{Method selection.} RQ-KMeans is simple, deterministic, and delivers strong AUC gains on every source we tested (Table~\ref{tab:fsq}) at several times smaller storage than RQ-FSQ; RQ-FSQ matches the dense float baseline at modestly higher storage by adding a reconstruction objective that preserves source geometry RQ-KMeans' nearest-centroid assignment can discard. Both produce the same $K$-level token sequence consumed by the HDE Module (Section~\ref{sec:hde}); the choice between them is a storage--fidelity trade-off rather than a binary preference.

\subsection{Multi-Source SID}
\label{sec:multisource-design}

We study three cross-domain user SID sources. All viewer SIDs are \emph{request-level}: constant across all sequence positions for a given user, keyed on user ID, and generated from data strictly available before the prediction timestamp. We use $K{=}3$ codes per source in all experiments.

A \textbf{Multi-Source SID} combines three cross-domain user embedding sources --- forming a progression of increasing behavioral activity richness --- into a single structured 9-code representation consumed by the HDE Module without architectural changes:
\begin{itemize}[leftmargin=*]
  \item $c_1$--$c_3$ \textbf{Activity-Tuned LLaMA SID}: embedding from a LLaMA-3.1-based model contrastively fine-tuned on cross-domain activity data with user profile as input. Encodes implicit behavioral intent even though input is profile text.
  \item $c_4$--$c_6$ \textbf{Profile Qwen SID}: embedding from a Qwen-based language model encoding user profile text (title, skills, summary). Encodes text semantics with no direct behavioral activity signal.
  \item $c_7$--$c_9$ \textbf{Feed Activity SID}: embedding aggregated from feed engagement signals over a 1-year trailing window. Provides the richest direct behavioral activity signal.
\end{itemize}
\paragraph{Backbone-based imputation for missing sources.}
Activity-Tuned LLaMA plays a second, distinct role beyond contributing $c_1$--$c_3$: it serves as the \textbf{backbone} for cross-source imputation when Profile Qwen or Feed Activity is unavailable for a user. We pick it for this role because (i)~it has the highest user coverage among the three sources --- so it is the source most often available when others are missing --- and (ii)~it is already activity-trained, making it a reasonable estimator of the missing source's content. Imputation preserves population coverage rather than silently zeroing out users with missing sources --- an important property in production deployments where no single source achieves 100\% coverage.

Concretely, let $\mathbf{u}$, $\mathbf{v}$, $\mathbf{w}$ denote the Activity-Tuned LLaMA, Profile Qwen, and Feed Activity embeddings respectively, with $\perp$ denoting a missing embedding. When $\mathbf{v}$ is observed, the Profile Qwen SID is $C_P = \text{RQ-KMeans}_P(\mathbf{v})$; when $\mathbf{v}$ is missing it is imputed from $\mathbf{u}$ through a dedicated residual VAE quantizer:
\begin{equation}
  \hat{C}_P = \begin{cases}
    \perp & \mathbf{v} = \perp,\ \mathbf{u} = \perp \\
    \text{RQ-VAE}_P\!\left(g(\mathbf{u})\right) & \mathbf{v} = \perp,\ \mathbf{u} \neq \perp \\
    C_P & \mathbf{v} \neq \perp,
  \end{cases}
  \label{eq:imputation}
\end{equation}
where $g$ is a linear projection from $\dim(\mathbf{u})$ to $\dim(\mathbf{v})$ and $\text{RQ-VAE}_P$ is a $K$-level residual VQ-VAE trained on users with both $\mathbf{u}$ and $\mathbf{v}$ present, with reconstruction target $\mathbf{v}$ and the standard objective:
\begin{equation}
\mathcal{L}_{\text{RQ-VAE}_P} = \Bigl\| \mathbf{v} - \sum_{l=1}^K \mathbf{e}^{(l)} \Bigr\|_2^2
  + \sum_{l=1}^{K} \Bigl(
      \bigl\| \operatorname{sg}[\mathbf{r}^{(l)}] - \mathbf{e}^{(l)} \bigr\|_2^2
    + \beta\, \bigl\| \mathbf{r}^{(l)} - \operatorname{sg}[\mathbf{e}^{(l)}] \bigr\|_2^2
  \Bigr),
  \label{eq:rqvae-imputation-loss}
\end{equation}
with $\mathbf{r}^{(l)}$ and $\mathbf{e}^{(l)}$ the residual and codebook embedding at level $l$, $\operatorname{sg}$ the stop-gradient operator, and $\beta$ the commitment loss coefficient. An independently trained $\text{RQ-VAE}_F$ is used for the Feed Activity source when $\mathbf{w}$ is missing. Each source uses its own codebook trained via RQ-KMeans: $C_A = \text{RQ-KMeans}_A(\mathbf{u})$, $C_P = \text{RQ-KMeans}_P(\mathbf{v})$, $C_F = \text{RQ-KMeans}_F(\mathbf{w})$.

\paragraph{Cascade fallback.} When the imputation backbone $\mathbf{u}$ is also missing (i.e., $\mathbf{v}=\mathbf{u}=\perp$ in Eq.~\ref{eq:imputation}, and symmetrically for $\mathbf{w}$), the affected source emits the padding code $c_k{=}0$ for all $K$ levels. The HDE Module maps padding codes to the zero embedding at every level (Eq.~\ref{eq:unigram}), so the user's representation cleanly degrades to the contribution of the remaining sources without spurious lookups.

\subsection{RQ-FSQ: Storage Footprint}
\label{sec:fsq}

An RQ-FSQ token sequence requires $\frac{K \cdot \lceil \log_2 C \rceil + D \cdot \lceil \log_2 L \rceil}{8} \approx 36$ bytes for our production configuration ($K{=}3$, $C{=}1024$, $D{=}64$, $L{=}16$). The reduction relative to the float32 source scales with the source dimensionality: $\sim 30\times$ for the lower-dimensional Feed Activity source, and $\sim 280\times$ for the higher-dimensional Activity-Tuned LLaMA source --- making RQ-FSQ particularly valuable for LLM-based encoders, with no custom preprocessing required at serving time.

\subsection{Hierarchical Discrete Embedding (HDE) Module}
\label{sec:hde}

The HDE Module encodes any $K$-level SID into a dense user embedding using \textbf{prefix n-gram sparse embedding tables} with hash-based lookups bounded by a memory cap $H^{\max}$. For a $K$-level SID $\mathbf{s} = (c_1, \ldots, c_K) \in \{0,\ldots,C\}^K$ (with $0$ denoting padding), each prefix $(c_1, \ldots, c_k)$ is hashed to an integer index and looked up in a dedicated trainable sparse table $\mathbf{W}_k$, yielding level-$k$ embedding $\mathbf{e}_k$. This is analogous to standard categorical embedding tables in recommendation systems, with prefix $k$-grams serving as the composite feature ID.

\paragraph{Level 1 --- prefix unigram.} Direct embedding lookup:
\begin{equation}
  \mathbf{e}_1 = \mathbf{W}_1[c_1], \quad \mathbf{W}_1 \in \mathbb{R}^{(C+1)\times d},
  \quad \mathbf{W}_1[0] = \mathbf{0}.
  \label{eq:unigram}
\end{equation}

\paragraph{Level $k \geq 2$ --- prefix $k$-gram.} Polynomial hashing over the prefix $(c_1, \ldots, c_k)$:
\begin{equation}
  \text{idx}_k = \Bigl[\sum_{j=1}^{k} (c_j{-}1) \cdot C^{k-j}\Bigr] \bmod H,
  \qquad
  \mathbf{e}_k = \mathbf{W}_k[\text{idx}_k],
  \quad \mathbf{W}_k \in \mathbb{R}^{H \times d}.
  \label{eq:kgram}
\end{equation}
All hash tables are capped at the same maximum size:
\begin{equation}
  H = \min\!\Bigl(\bigl\lfloor C^K / \alpha \bigr\rfloor,\; H^{\max}\Bigr),
  \label{eq:hashsize}
\end{equation}
where $\alpha$ is a compression factor and $H^{\max}$ caps memory. Level embeddings are summed to produce the user embedding $\mathbf{e} = \sum_{k=1}^{K} \mathbf{e}_k$. All tables $\{\mathbf{W}_k\}_{k=1}^{K}$ are initialized randomly and trained jointly under the CTR objective.

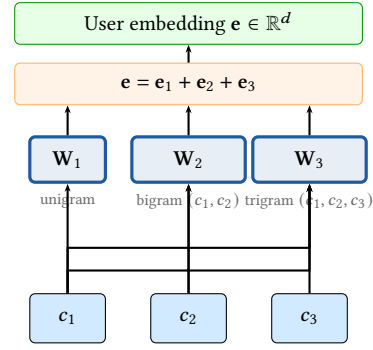
\begin{figure}[t]
\centering
\begin{tikzpicture}[
  font=\small,
  tok/.style={rectangle, rounded corners=2pt, draw, minimum width=1.0cm, minimum height=0.65cm, text centered, inner sep=3pt},
  tbl/.style={rectangle, rounded corners=2pt, draw=accentblue, fill=accentblue!10, very thick, minimum width=1.5cm, minimum height=0.60cm, text centered, inner sep=3pt},
  arr/.style={-{Stealth[length=3pt]}, thick},
  darr/.style={-{Stealth[length=3pt]}, thick, color=mutedgray},
]

\node[tok, fill=activitycol!60]  (c1) at (0.0, 0) {$c_1$};
\node[tok, fill=activitycol!60]  (c2) at (1.6, 0) {$c_2$};
\node[tok, fill=activitycol!60]  (c3) at (3.2, 0) {$c_3$};

\node[tbl, minimum width=1.0cm] (w1) at (0.0, 2.1) {$\mathbf{W}_1$};
\node[font=\scriptsize, color=mutedgray] at (0.0, 1.55) {unigram};

\node[tbl] (w2) at (1.6, 2.1) {$\mathbf{W}_2$};
\node[font=\scriptsize, color=mutedgray] at (1.6, 1.55) {bigram $(c_1,c_2)$};

\node[tbl] (w3) at (3.2, 2.1) {$\mathbf{W}_3$};
\node[font=\scriptsize, color=mutedgray] at (3.2, 1.55) {trigram $(c_1,c_2,c_3)$};

\draw[arr] (c1.north) -- (w1.south);

\draw[arr] (c1.north) -- ++(0, 0.30) -| (w2.south);
\draw[arr] (c2.north) -- (w2.south);

\draw[arr] (c1.north) -- ++(0, 0.60) -| (w3.south);
\draw[arr] (c2.north) -- ++(0, 0.30) -| (w3.south);
\draw[arr] (c3.north) -- (w3.south);

\node[rectangle, draw=orange!60, fill=orange!10, rounded corners=2pt,
      minimum width=4.6cm, minimum height=0.55cm, text centered]
      (agg) at (1.6, 3.10)
      {$\mathbf{e} = \mathbf{e}_1 + \mathbf{e}_2 + \mathbf{e}_3$};

\draw[arr] (w1.north) -- (w1.north |- agg.south);
\draw[arr] (w2.north) -- (w2.north |- agg.south);
\draw[arr] (w3.north) -- (w3.north |- agg.south);

\node[rectangle, draw=green!60!black, fill=green!10, rounded corners=2pt,
      minimum width=4.6cm, minimum height=0.55cm, text centered]
      (out) at (1.6, 3.90)
      {User embedding $\mathbf{e} \in \mathbb{R}^d$};
\draw[arr] (agg) -- (out);

\end{tikzpicture}
\caption{\textbf{Hierarchical Discrete Embedding (HDE) Module} illustrated for a single-source $K{=}3$ SID.
  \textbf{Level 1} looks up the prefix unigram $c_1$ in a direct table $\mathbf{W}_1$.
  \textbf{Level 2} hashes the prefix bigram $(c_1, c_2)$ into table $\mathbf{W}_2$.
  \textbf{Level 3} hashes the prefix trigram $(c_1, c_2, c_3)$ into table $\mathbf{W}_3$ (note arrows from all three tokens).
  All tables use hash size $H = \min(\lfloor C^K/\alpha \rfloor, H^{\max})$ and are trained end-to-end under the CTR objective; level embeddings are summed to produce $\mathbf{e}$.
  In the Multi-Source SID setting, the same module is applied independently to each source's 3-code block ($c_1$--$c_3$, $c_4$--$c_6$, $c_7$--$c_9$) and all embeddings are summed (Eq.~\ref{eq:multisource-hde}).}
\label{fig:hde}
\end{figure}

\paragraph{Multi-Source SID.} For the Multi-Source SID (Section~\ref{sec:multisource-design}), prefix n-gram tables are applied independently to each source's $K$-code block and all embeddings are summed:
\begin{equation}
  \mathbf{e}^{\mathrm{user}} = \sum_{s=1}^{S} \mathrm{HDE}\!\bigl(\mathbf{s}^{(s)}\bigr),
  \label{eq:multisource-hde}
\end{equation}
with $\mathbf{s}^{(1)}{=}(c_1,c_2,c_3)$, $\mathbf{s}^{(2)}{=}(c_4,c_5,c_6)$, $\mathbf{s}^{(3)}{=}(c_7,c_8,c_9)$, yielding $S{\times}K{=}9$ table lookups (Figure~\ref{fig:hde}).

\subsection{Integration into the Ads Ranking Model}
\label{sec:integration}

User SIDs are \emph{request-level}: the same $K$-token sequence is broadcast across all $L$ positions in the interaction sequence for a given user. The input representation at event $e_t$ is:
\begin{equation}
  \mathbf{h}_t = \text{LayerNorm}\!\Bigl(
    \sum_{f \in \mathcal{F}} \mathbf{e}_t^{(f)} + \mathbf{e}^{\mathrm{user}}
  \Bigr),
  \label{eq:fusion}
\end{equation}
where $\mathcal{F}$ is the standard feature set and $\mathbf{e}^{\mathrm{user}} = \mathrm{HDE}(\mathbf{s}_{\mathrm{user}})$ is the user embedding from the HDE Module, constant across positions.

\paragraph{Learning rates.} HDE embedding tables use a higher learning rate ($\eta_{\mathrm{HDE}} = 0.02$) than Transformer weights ($\eta_{\mathrm{TR}} = 4\times10^{-4}$), following established practice for embedding-heavy ranking systems~\citep{din2018,dcnv2_2021}.

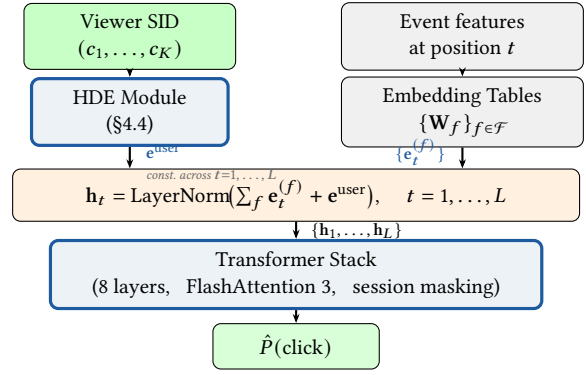
\begin{figure*}[t]
\centering
\begin{tikzpicture}[
  font=\small,
  box/.style={rectangle, rounded corners=3pt, draw, minimum height=0.60cm,
              text centered, inner sep=4pt},
  boldbox/.style={rectangle, rounded corners=3pt, draw=accentblue, very thick,
                  minimum height=0.60cm, text centered, inner sep=4pt, fill=accentblue!8},
  arr/.style={-{Stealth[length=3.5pt]}, thick},
]

\node[box, fill=green!22, minimum width=2.8cm, align=center]
  (sid) at (2.8, 2.8) {Viewer SID\\$(c_1,\ldots,c_K)$};

\node[boldbox, minimum width=2.6cm, align=center]
  (hde) at (2.8, 1.8) {HDE Module\\(§\ref{sec:hde})};

\node[box, fill=gray!12, minimum width=3.2cm, align=center]
  (events) at (7.2, 2.8) {Event features\\at position $t$};

\node[box, fill=gray!12, minimum width=3.2cm, align=center]
  (embtables) at (7.2, 1.8) {Embedding Tables\\$\{\mathbf{W}_f\}_{f\in\mathcal{F}}$};

\node[box, fill=orange!12, minimum width=7.5cm, align=center]
  (fusion) at (5.0, 0.7)
  {$\mathbf{h}_t = \text{LayerNorm}\!\bigl(\textstyle\sum_{f} \mathbf{e}_t^{(f)} + \mathbf{e}^{\mathrm{user}}\bigr)$, \quad $t = 1,\ldots,L$};

\node[boldbox, minimum width=6.5cm, minimum height=0.90cm, align=center]
  (transformer) at (5.0, -0.35)
  {Transformer Stack\\(8 layers, \; FlashAttention~3, \; session masking)};

\node[box, fill=green!15, minimum width=2.2cm]
  (pclick) at (5.0, -1.30) {$\hat{P}(\text{click})$};

\draw[arr] (sid) -- (hde);
\draw[arr] (events) -- (embtables);
\draw[arr] (hde.south) -- (hde.south |- fusion.north);
\draw[arr] (embtables.south) -- (embtables.south |- fusion.north);
\draw[arr] (fusion)
  -- node[right=2pt, font=\scriptsize]{$\{\mathbf{h}_1,\ldots,\mathbf{h}_L\}$}
  (transformer);
\draw[arr] (transformer) -- (pclick);

\node[font=\scriptsize, color=accentblue, right=3pt] at (2.8, 1.27)
  {$\mathbf{e}^{\mathrm{user}}$};
\node[font=\tiny\itshape, color=mutedgray, right=3pt] at (2.8, 0.97)
  {const.\ across $t{=}1,\ldots,L$};
\node[font=\scriptsize, color=accentblue, left=3pt] at (7.2, 1.27)
  {$\{\mathbf{e}_t^{(f)}\}$};

\end{tikzpicture}
\caption{\textbf{Integration of the HDE Module into the Ads Ranking Model.}
  \textbf{Left}: The viewer SID $(c_1,\ldots,c_K)$ is encoded once per request by the HDE Module (Section~\ref{sec:hde}), producing $\mathbf{e}^{\mathrm{user}}$ --- a dense vector \emph{constant across all $L$ positions} in the sequence.
  \textbf{Right}: Per-event features at position $t$ are encoded through standard embedding tables.
  \textbf{Fusion} (Eq.~\ref{eq:fusion}): At every position $t$, the user embedding is summed with the event-level embeddings before a shared LayerNorm, yielding input tokens $\{\mathbf{h}_t\}$.
  The resulting sequence is processed by the Transformer stack and scored by the CTR head.
  All HDE embedding tables are trained end-to-end under the CTR objective.}
\label{fig:integration}
\end{figure*}

\paragraph{Data pipeline.} User SID tokens are materialized offline, normalized to exactly $K$ levels during data loading, and broadcast to $[B, L, K]$ integer tensors during batch collation, ensuring strict train-serve schema parity. Figure~\ref{fig:integration} illustrates the integration.

\section{Experiments}

Our experiments answer five research questions:
\begin{itemize}[leftmargin=*,itemsep=1pt]
  \item \textbf{RQ1}: Can cross-domain viewer SIDs improve ads CTR ranking over the no-SID production baseline?
  \item \textbf{RQ2}: Does the AUC gain from a cross-domain SID scale with the amount of behavioral activity encoded in its source embedding?
  \item \textbf{RQ3}: Does the structured Multi-Source SID with backbone-based imputation beat na\"{i}vely summing independent single-source SIDs?
  \item \textbf{RQ4}: Does RQ-FSQ match the AUC of the dense source embedding across heterogeneous pre-trained sources?
  \item \textbf{RQ5}: Are cross-domain viewer SIDs most valuable for users with sparse ad interaction history (cold-start)?
\end{itemize}

\subsection{Setup}

\paragraph{Data.} Production ads logs from a large-scale industrial recommendation platform; 60 days of data for training and the subsequent 1 day for evaluation. All results are reported as relative AUC gains over the no-SID baseline; absolute values are withheld per confidentiality policy.

\paragraph{Model.} The production ads click prediction ranking model described in Section~\ref{sec:baseline}. User SID default: codebook $C{=}1024$, levels $K{=}3$, hash size $H = \min(\lfloor C^K/\alpha \rfloor, H^{\max})$, embedding dimension $D=64$. For RQ-VAE, commitment loss coefficient $\beta$ is 0.25.

\paragraph{Metric.} AUC relative to the no-SID baseline. At industrial production scale, $+0.1\%$ offline AUC reliably corresponds to measurable online CTR impact; this correspondence has been validated on the same production system across multiple prior deployments~\citep{cadet2026}.

\paragraph{Controls.} Random seed, batch size, optimizer schedule, and model depth are held fixed across all variants to isolate SID effects.

\paragraph{Hyperparameter sensitivity.} We selected $K{=}3$, $C{=}1024$, and $D{=}64$ based on internal sweeps on pilot configurations: smaller $K$ underfit, and larger $K$ or $C$ yielded no measurable improvement at our deployment scale. The hash-table cap $H^{\max}$ is a memory budget that bounds embedding-table size to keep downstream serving costs manageable, with no observed quality cost in our pilot studies. For RQ-FSQ, we use $L{=}16$ levels per dimension (4 bits), chosen to keep the FSQ branch compact while retaining enough per-dimension resolution to preserve fine structure.

\paragraph{Serving cost.} The HDE Module adds negligible inference latency over the no-SID baseline: all lookups are local memory accesses on standard categorical embedding tables, and per-request user SIDs are precomputed offline (Section~\ref{sec:integration}). Training cost is essentially unchanged because the HDE tables share the same SGD step as the rest of the model.

\subsection{Main Results (RQ1--RQ3)}

Table~\ref{tab:ablation} consolidates all SID results. The upper block covers single-source SIDs (RQ1, RQ2); the lower block covers multi-source combinations (RQ3). All methods use the same production Ads ranking model and HDE Module with $K{=}3$ codes per source.

\begin{table*}[t]
\centering
\caption{Cross-domain user SID results (RQ-KMeans quantization, $K{=}3$ codes per source). $\Delta$AUC relative to no-SID baseline. \emph{Upper block}: single-source SIDs ordered by behavioral activity richness. \emph{Lower block}: multi-source combinations. The RQ-FSQ comparison against the same sources appears in Table~\ref{tab:fsq}.}
\label{tab:ablation}
\begin{tabular}{llc}
\toprule
\textbf{Method} & \textbf{Description} & $\boldsymbol{\Delta}$ \textbf{AUC} \\
\midrule
No SID & Reference & --- \\
\midrule
\multicolumn{3}{l}{\textit{Single-source SIDs (behavioral activity richness $\nearrow$)}} \\
\midrule
Profile Qwen SID         & Text semantics only, no behavioral signal    & $+0.036\%$ \\
Activity-Tuned LLaMA SID & Activity-trained, profile-prompted           & $+0.107\%$ \\
Feed Activity SID        & Direct behavioral signals (1-yr agg.)        & $+0.213\%$ \\
\midrule
\multicolumn{3}{l}{\textit{Multi-source combinations ($K{=}3$ per source, 9 total codes)}} \\
\midrule
Independent combination  & 3 SIDs, separately indexed and summed        & $+0.260\%$ \\
\textbf{Multi-Source SID} & Structured 9-code, Activity-Tuned LLaMA backbone & $\mathbf{+0.296\%}$ \\
\bottomrule
\end{tabular}
\end{table*}

\subsection{Behavioral Activity Richness Governs Transfer Quality (RQ2)}
\label{sec:rq2}

The upper block of Table~\ref{tab:ablation} reveals a systematic relationship: the downstream AUC gain scales monotonically with the amount of behavioral activity encoded in the source representation. Profile Qwen SID (text semantics, no behavioral signal) yields $+0.036\%$. Activity-Tuned LLaMA SID, fine-tuned on cross-domain activity data, yields $+0.107\%$ --- the implicit behavioral signal embedded during its training provides substantial additional value. Feed Activity SID, directly aggregating 1-year of engagement signals, yields $+0.213\%$.

We term this the \textbf{behavioral activity richness principle}: for cross-domain SID transfer, the quality of the transferred signal is determined by how much behavioral activity is encoded in the source representation, regardless of whether that activity is encoded directly (Feed Activity) or implicitly through an activity-trained model (Activity-Tuned LLaMA). Profile attributes alone provide a comparatively weak signal ($+0.036\%$), consistent with the intuition that \emph{what a user does} is more predictive of ad engagement than \emph{how their profile is described}.

\paragraph{Mechanism.} Feed activity captures high-frequency, evolving behavioral intent absent from both ads interaction logs and user profile descriptions. Discretizing these signals and fine-tuning end-to-end under the CTR loss performs \emph{implicit domain adaptation}: the gradient re-specializes the embedding tables to map feed behavioral clusters toward ad engagement probability, with no explicit cross-domain alignment objective.

\subsection{Multi-Source SID (RQ3)}
\label{sec:multisource}

The lower block of Table~\ref{tab:ablation} shows the multi-source results. Independent combination --- summing three separately-indexed single-source SIDs with no structural coordination --- already yields $+0.260\%$, confirming that all three sources contribute complementary signal. The structured Multi-Source SID (Section~\ref{sec:multisource-design}) achieves $+0.296\%$, a $+0.036\%$ gain over independent combination at identical parameter budget.

The gain comes from two factors. First, per-source prefix n-gram tables avoid cross-source hash collisions that arise when independently-indexed tables are summed without source partitioning. Second, the Activity-Tuned LLaMA backbone fallback for missing Profile Qwen codes preserves population coverage rather than silently zeroing out the affected users.

\subsection{RQ-FSQ Across Pre-Trained Embeddings (RQ4)}
\label{sec:fsq-exp}

We evaluate RQ-FSQ on two heterogeneous pre-trained sources --- a lower-dimensional Feed Activity embedding and a higher-dimensional Activity-Tuned LLaMA embedding --- against the dense float baseline, RQ-KMeans, and FSQ alone.

\begin{table*}[t]
\centering
\caption{RQ-FSQ on two heterogeneous pre-trained embedding sources.
$\boldsymbol{\Delta}$\textbf{AUC} is measured against the \emph{no-SID baseline} --- the production ranking model with no member SID input.
\textbf{Storage} is the per-user serialization footprint relative to the \emph{raw float32 embedding} of the same source.
RQ-FSQ matches the dense embedding's AUC on both sources at $30$--$280{\times}$ smaller storage.}
\label{tab:fsq}
\setlength{\tabcolsep}{8pt}
\renewcommand{\arraystretch}{1.15}
\begin{tabular}{lcccc}
\toprule
& \multicolumn{2}{c}{\textbf{Feed Activity}} & \multicolumn{2}{c}{\textbf{Activity-Tuned LLaMA}} \\
& \multicolumn{2}{c}{\footnotesize\itshape lower-dimensional source} & \multicolumn{2}{c}{\footnotesize\itshape higher-dimensional source} \\
\cmidrule(lr){2-3}\cmidrule(lr){4-5}
\textbf{Method} & \textbf{Storage} & $\boldsymbol{\Delta}$\textbf{AUC} & \textbf{Storage} & $\boldsymbol{\Delta}$\textbf{AUC} \\
\midrule
Raw float embedding (dense) & $1{\times}$              & $+0.349\%$          & $1{\times}$               & $+0.264\%$          \\
RQ-KMeans                    & ${\sim}\,0.004{\times}$   & $+0.213\%$          & ${\sim}\,0.0004{\times}$   & $+0.107\%$          \\
FSQ                          & ${\sim}\,0.03{\times}$    & $+0.343\%$          & ${\sim}\,0.003{\times}$    & $+0.248\%$          \\
\midrule
\textbf{RQ-FSQ (ours)}       & $\boldsymbol{\sim\,0.03{\times}}$ & $\mathbf{+0.351\%}$ & $\boldsymbol{\sim\,0.003{\times}}$ & $\mathbf{+0.265\%}$ \\
\bottomrule
\end{tabular}
\end{table*}

The pattern is consistent across both sources. RQ-KMeans reduces storage substantially but loses AUC because deterministic centroid assignment does not optimize a reconstruction objective. FSQ alone recovers most of the gap by preserving per-dimension structure. RQ-FSQ matches or slightly exceeds the dense float baseline on both sources, because the reconstruction loss in the RQ-VAE branch preserves global embedding geometry while the FSQ branch preserves per-dimension fine structure. The storage benefit grows with source dimensionality: $\sim 30\times$ for the lower-dimensional Feed Activity source, $\sim 280\times$ for the higher-dimensional Activity-Tuned LLaMA source --- making RQ-FSQ particularly valuable for higher-dimensional LLM-based encoders, where dense float storage is prohibitive at production scale.

\subsection{Cross-Domain User SIDs and User Cold-Start (RQ5)}
\label{sec:coldstart}

The cold-start hypothesis --- that cross-domain viewer SIDs are most valuable for users with sparse ad interaction history --- is directly testable by stratifying by user activity level. We partition the validation set into three segments by number of distinct ad impressions in the trailing history: \emph{most cold-start} (near-zero ad history; bottom 8\% of users by trailing-history size), \emph{infrequent} (the intermediate 64\%), and \emph{frequent} (top 28\% of users by trailing-history size).

\begin{table}[t]
\centering
\caption{Feed Activity SID gain by user activity segment. Segments are defined by the number of distinct ad impressions in the trailing history window.}
\label{tab:coldstart}
\begin{tabular}{lc}
\toprule
\textbf{User segment} & $\boldsymbol{\Delta}$ \textbf{AUC (Feed Activity SID)} \\
\midrule
Most cold-start & $+1.522\%$ \\
Infrequent      & $+0.874\%$ \\
Frequent        & $+0.131\%$ \\
\midrule
Overall         & $+0.213\%$ \\
\bottomrule
\end{tabular}
\end{table}

The pattern directly validates the cross-domain transfer mechanism: gains scale monotonically with sparsity. Most cold-start users benefit the most ($+1.522\%$), where the ads history is near-empty and the cross-domain feed activity SID provides the \emph{primary} behavioral evidence available to the model. Infrequent users gain $+0.874\%$, and frequent users --- who already have rich first-party ads signals --- gain $+0.131\%$, with feed activity SIDs serving as a complementary source. Feed activity SIDs effectively act as a \emph{behavioral cold-start bridge}, delivering the largest per-impression benefit precisely where the existing model is weakest.

\subsection{Public-Data Validation}
\label{sec:public}

To confirm the RQ-FSQ $>$ RQ-KMeans ordering generalizes beyond our production data, we replicate the comparison on the public MovieLens-100K benchmark. Movie text (title $+$ genres) is encoded with the open-weight \texttt{all-MiniLM-L6-v2} sentence encoder; user embeddings are the mean of liked-movie embeddings, the public analog of cross-domain behavioral aggregation. We quantize user embeddings via RQ-KMeans, FSQ, and RQ-FSQ ($K{=}3, C{=}256, L{=}7$, scaled down for the smaller dataset) and train a small MLP to predict whether a held-out user--movie pair has rating $\geq 4$; RQ-FSQ is warm-started from the trained RQ-KMeans weights, mirroring deployment of the FSQ residual on top of an existing RQ-KMeans system. Table~\ref{tab:public} confirms RQ-FSQ achieves the highest AUC, exceeding both RQ-KMeans and the dense user embedding --- matching Table~\ref{tab:fsq} on a public benchmark.

\begin{table}[t]
\centering
\caption{Public replication on MovieLens-100K. $\Delta$AUC vs.\ a no-SID (movie-only) baseline.}
\label{tab:public}
\setlength{\tabcolsep}{6pt}
\begin{tabular}{lcc}
\toprule
\textbf{Method} & \textbf{AUC} & $\boldsymbol{\Delta}$\textbf{AUC} \\
\midrule
No SID (baseline)      & $0.7689$ & --- \\
Dense user emb         & $0.8078$ & $+3.89\%$ \\
RQ-KMeans              & $0.8215$ & $+5.26\%$ \\
\textbf{RQ-FSQ (ours)} & $\mathbf{0.8343}$ & $\mathbf{+6.54\%}$ \\
\bottomrule
\end{tabular}
\end{table}

\section{Discussion}

\paragraph{Behavioral activity richness as a general design principle.}
The monotonic progression from profile text ($+0.036\%$) to activity-trained embeddings ($+0.107\%$) to direct behavioral aggregations ($+0.213\%$) gives a concrete source-selection rule: prefer sources encoding recent behavioral signals over static attribute descriptions. This extends the established finding that behavioral features dominate demographics for engagement prediction~\citep{din2018,dien2019} to the cross-domain setting --- behavioral richness in the source survives the discrete bottleneck and still drives downstream gains after quantization.

\paragraph{RQ-FSQ as a general-purpose discretizer for pre-trained embeddings.}
RQ-FSQ matches or slightly exceeds dense-float AUC across two heterogeneous pre-trained sources because the branches are complementary: FSQ preserves per-dimension fine structure RQ-VAE can compress away; RQ-VAE preserves global geometry FSQ ignores. The design should generalize to any pipeline needing a compact discrete code over a pre-trained vector --- item embeddings, multimodal encoders, LLM-based representations.

\paragraph{Cross-domain transfer via discrete representations.}
Our results provide the first empirical evidence that viewer SIDs from a different domain (organic feed) are effective input features for ads CTR, and can outperform within-domain profile SIDs. The key enabler is the combination of \emph{discrete bottleneck} and \emph{end-to-end fine-tuning}: quantization strips domain-specific embedding geometry while preserving semantic cluster structure; the CTR gradient re-specializes the embedding tables toward the target task. This is qualitatively analogous to transfer learning via pre-training and fine-tuning, but operating in the discrete token space.

\paragraph{Generalizable artifacts.}
Our methodological contributions --- RQ-FSQ, the HDE Module, Multi-Source SID with backbone-based imputation, and the behavioral activity richness principle --- depend only on pre-trained user embeddings and a downstream ranking objective, applying to any system meeting those conditions. Our three sources (Profile Qwen, Activity-Tuned LLaMA, Feed Activity) are platform-specific, but each maps to a category readily available in most large recommendation systems --- a profile-text encoder, an activity-tuned user encoder, and a behavioral aggregation --- reproducible with open-weight analogues. The public-data replication on MovieLens-100K (Section~\ref{sec:public}) confirms the RQ-FSQ $>$ RQ-KMeans ordering on an open benchmark.

\paragraph{Broader impacts.}
The primary effect of cross-domain user SIDs in deployment is improved cold-start ranking quality, which reduces low-relevance impressions for users with sparse ad-domain history. The same mechanism increases the granularity at which inferred user interests can be associated across domains, so production use should remain subject to the standard consent and privacy controls already governing behavioral feature pipelines. The discrete bottleneck of SIDs is privacy-favorable relative to dense vector exposure: only $K$ low-bit codes are materialized per user, limiting the resolution at which raw activity is preserved.

\section{Conclusion}

We presented \textbf{RQ-FSQ}, a quantization method that pairs per-dimension scalar quantization with residual VAE quantization, matching or slightly exceeding dense-float AUC at $\sim 30\times$--$280\times$ storage reduction across two heterogeneous embedding sources. We established the \textbf{behavioral activity richness principle} --- cross-domain transfer scales monotonically with the behavioral signal encoded in the source ($+0.036\%$/$+0.107\%$/$+0.213\%$) --- and introduced the \textbf{HDE Module} and \textbf{Multi-Source SID} ($+0.296\%$ AUC), deployable without architectural changes. Cold-start segment analysis confirms the transfer mechanism, with gains of $+1.522\%$ on the most cold-start users --- cross-domain SIDs serve as a behavioral bridge precisely where ads-domain history fails.

\paragraph{Outlook.} The same SID interface extends naturally to other ranking surfaces and multi-task heads beyond CTR --- the discrete-token bottleneck enables a single unified user representation across the production stack, with shared HDE tables serving as the common substrate.


\bibliographystyle{unsrtnat}
\bibliography{refs}

\end{document}